\documentclass[11pt,preprint,showpacs,preprintnumbers,amsmath,amssymb]{revtex4}

\usepackage[breaklinks,colorlinks,citecolor=blue,urlcolor=blue,linkcolor=blue]{hyperref}
\usepackage{graphicx}
\usepackage{dcolumn}
\usepackage{bm}
\usepackage{multirow}
\usepackage{float}
\usepackage{array}
\usepackage{booktabs}
\usepackage{caption}
\usepackage{txfonts}
\usepackage{array}
\usepackage{setspace}
\usepackage{amssymb}
\usepackage{CJK}
\usepackage{caption}
\usepackage{rotating}
\usepackage{subfigure}
\usepackage{relsize}
\usepackage{diagbox}
\usepackage{epstopdf}
\usepackage{pifont}
\usepackage{rotating}

\begin{document}

\title{Investigating $S$-wave bound states composed of two pseudoscalar mesons}

\author{Zi-Long Li}
\affiliation{\scriptsize{Physics Department, Ningbo University, Zhejiang 315211, China}}

\author{Xing-Yue Tan}
\affiliation{\scriptsize{Physics Department, Ningbo University, Zhejiang 315211, China}}

\author{Zhu-Feng Zhang}
\affiliation{\scriptsize{Physics Department, Ningbo University, Zhejiang 315211, China}}

\author{Zhen-Yang Wang \footnote{Corresponding author, e-mail: wangzhenyang@nbu.edu.cn}}
\affiliation{\scriptsize{Physics Department, Ningbo University, Zhejiang 315211, China}}

\author{Xin-Heng Guo \footnote{Corresponding author, e-mail: xhguo@bnu.edu.cn}}
\affiliation{\scriptsize{College of Nuclear Science and Technology, Beijing Normal University, Beijing 100875, China}}

\date{\today}

\begin{abstract}
In this work, we systematically investigate the two-pseudoscalar meson systems with the Bethe-Salpeter equation in the ladder and instantaneous approximations. By solving the Bethe-Salpeter equation numerically with the kernel containing the one-particle exchange diagrams, we find that the $K\bar{K}$, $DK$, $B\bar{K}$, $D\bar{D}$, $B\bar{B}$, $BD$, $D\bar{K}$, $BK$, and $B\bar{D}$ systems with $I=0$ can exist as bound states. We also study the contributions from heavy meson ($J/\psi$ and $\Upsilon$) exchanges, and we find that the contribution from heavy meson exchange can not be ignored.

\end{abstract}

\pacs{***}

\maketitle

\section{Introduction}
Quantum chromodynamics (QCD) is the theory of the strong interaction between quarks mediated by gluons, which are color-charged \cite{Han:1965pf}. In principle, QCD allows complex quark and gluon compositions of hadrons, like multiquark hadrons, hadronic molecules, hybrid hadrons, and glueballs which are nonstandard hadronic particles. Most of these nonstandard hadrons have unusual masses, decay widths, etc., which cannot be given a satisfactory explanation by the traditional quark model. Up to now, more than thirty non-$q\bar{q}$ state candidates in light and heavy sectors have been reported experimentally \cite{ParticleDataGroup:2020ssz}. These resonances are crucial for the deep understanding of the hadron spectroscopy and the nonperturbative nature and spontaneous chiral symmetry breaking of QCD.

In the low-lying scalar meson sector, $f_0(980)$ \cite{Astier:1967zz}, $a_0(980)$ \cite{Ammar:1968zur} and $D^\ast_{s0}(2317)$ \cite{BaBar:2003oey} have nonexotic $J^P (=0^+)$ quantum numbers. However, their masses are much lower than the quark model expectation for the corresponding $P$-wave $q\bar{q}$ states \cite{Godfrey:2015dva,Olsen:2017bmm}. Their natures are still under debate in spite of the efforts during the past several decades. Since their masses are near the threshold of the constituent particles and have spin-parity quantum numbers corresponding to the $S$-wave combinations of the constituent particles, one would naturally identify them as hadronic molecules, which are analogs of nuclei. $f_0(980)$ and/or $a_0(980)$ could be $K\bar{K}$ molecules \cite{Weinstein:1982gc,Weinstein:1983gd,Weinstein:1990gu,Oller:2002na,Baru:2003qq,Dai:2014zta,Ahmed:2020kmp,Wang:2022vga} and $D^\ast_{s0}(2317)$ could be a $DK$ molecule \cite{Szczepaniak:2003vy,Hofmann:2003je,Guo:2006fu,Flynn:2007ki,Xie:2010zza,Guo:2015dha,Du:2017ttu,Wu:2019vsy,Kong:2021ohg,Huang:2021fdt,Zhang:2006ix} . Such a picture leads to the results consistent with the experiments. Besides these particles, the possible $S$-wave bound states of $\bar{B}{K}$, $D\bar{D}$, $B\bar{B}$, $BD$, $\bar{B}\bar{K}$, $DD$, $\bar{B}\bar{B}$, and $\bar{B}D$ systems have not been observed experimentally.

On the theoretical aspect, the authors in Ref. \cite{Zhang:2006ix} systematically study the possible $S$-wave bound state of two pseudoscalar mesons by the nonrelativistic Schr$\ddot{o}$dinger (NRS) equation. Ref. \cite{Guo:2006fu} predicted the existence of a $B\bar{K}$ bound state $B_{s0}^\ast$ with a mass of 5.725 $\pm$ 0.039 GeV based on the heavy chiral unitary approach. Subsequently, Refs. \cite{Faessler:2008vc} and \cite{Feng:2011zzb} confirmed the existence of $B_{s0}^\ast$ in the $B\bar{K}$ bound state scenario and further studied the decay widths of its possible decay channels. Recently, Kong et al. \cite{Kong:2021ohg} systematically investigated $DK$/$\bar{B}K$ and $\bar{D}K$/$BK$ systems in a quasipotential Bethe-Salpter equation (qBSE) approach by considering the light meson exchange potential, and found only the isoscalar systems can exist as molecular states. However, the mass of $X(5568)$ reported by the D0 collaboration \cite{D0:2016mwd} is too far below the $BK$ threshold to be a $BK$ molecule \cite{Xiao:2016mho,Agaev:2016urs,Chen:2016npt,Chen:2016ypj,Lu:2016kxm,Wang:2018jsr}. In Ref. \cite{Liu:2008mi},  the authors studied the $S$-wave $D\bar{D}$, $BD$ and $B\bar{B}$ systems in chiral SU(3) quark model (QM), their calculation favors the existence of the isoscalar $B\bar{B}$ molecule but the existence of isovector $D\bar{D}$ and $BD$ molecules is disfavored. In Ref. \cite{Ohkoda:2012hv}, $D^{(\ast)}D^{(\ast)}$ and $B^{(\ast)}B^{(\ast)}$ molecular states were studied by solving the coupled channel Schr$\mathrm{\ddot{o}}$dinger (CCS) equations, only the $I(J^P)=1(0^+)$ $BB$ can be a bound state in the $PP$ ($P=D, B$) system because of the the kinetic term is suppressed in the bottom sector and the effect of channel couplings becomes more important. With the qBSE approach \cite{Ding:2020dio}, the existence of $D\bar{D}$ and $B\bar{B}$ molecular states with $I(J^P)=0(0^+)$ were predicted, yet no bound state was produced from the $DD$ and $\bar{B}\bar{B}$ interaction. In Ref. \cite{Gamermann:2006nm}, a new hidden charm resonance with mass 3.7 GeV was predicted within the coupled channel unitary approach. Later the $D\bar{D}$ bound state was searched in several processes, such as $B \rightarrow D \bar{D} K$ \cite{Dai:2015bcc}, $\psi (3770) \rightarrow \gamma D^0 {\bar{D}}^0$ \cite{Dai:2020yfu}, and $ \gamma\gamma \rightarrow D\bar D$ \cite{Wang:2020elp,Deineka:2021aeu}. There are some differences in the results of different methods. Therefore, more efforts are needed to investigate the possible $S$-wave bound state composed of two pseudoscalar mesons.

In the present paper, we will systematically investigate whether the $S$-wave bound states of two-pseudoscalar meson systems exist in the Bethe-Salpeter (BS) approach (in the ladder approximation and the instantaneous approximation for the kernel). For the doubly heavy pseudoscalar meson systems, we will not only consider the interaction through exchanged light mesons ($\rho$, $\omega$ and $\sigma$), but also the contribution of heavy vector mesons ($J/\psi$ or $\Upsilon$). As studied in \cite{Aceti:2014kja,Aceti:2014uea}, in spite of the large mass of the $J/\psi$, which suppresses the propagator of the exchanged $J/\psi$, it was found that the interaction could bind the $D^\ast\bar{D}^\ast$ and $D\bar{D}^\ast$ systems. Similarly, Refs. \cite{Ding:2020dio,Ding:2021igr} also found the contribution from heavy meson exchange ($J/\psi$ or $\Upsilon$) is very important to form a molecular state, especially in the systems with the contributions from $\rho$ and $\omega$ canceling each other.

The remainder of this paper is organized as follows. In Sec. \ref{sect-BS-PP}, we discuss the BS equation for the two-pseudoscalar meson systems and establish the BS equation for this system. This equation is solved numerically and the numerical results of the two-pseudoscalar meson systems are presented in Sec. \ref{Num}. In the last section, we give a summary.

\section{The bethe-salpeter formalism for the two-pseudoscalar meson system}
\label{sect-BS-PP}
The BS wave function for the bound state $|P\rangle$ composed of two pseudoscalar mesons have the following form:
\begin{equation}
  \chi\left(x_1,x_2,P\right) = \langle0|T\mathcal{P}_1(x_1)\mathcal{P}_2(x_2)|P\rangle,
\end{equation}
where $\mathcal{P}_1(x_1)$ and $\mathcal{P}_2(x_2)$ are the field operators of the two constituent particles at space coordinates $x_1$ and $x_2$, respectively. The BS wave function in momentum space is defined as
\begin{equation}\label{PP-momentum-BS-function}
 \chi_P(x_1,x_2,P) = e^{-iPX}\int\frac{d^4p}{(2\pi)^4}e^{-ipx}\chi_P(p),
\end{equation}
where $p$ represents the relative momentum of the two constituent particles and $p= \lambda_2 p_1-\lambda_1 p_2$ ($p_1=\lambda_1P+p$,\quad $p_2=\lambda_2P-p$) with $\lambda_1 = m_1/(m_1 + m_2)$ and $\lambda_2 = m_2/(m_1 + m_2)$, $p_{1(2)}$ and $m_{1(2)}$ represent the momentum and mass of the constituent particle, respectively.

The BS wave function $\chi_P(p)$ satisfies the following BS equation:
\begin{equation}\label{BS-equation}
  \chi_{P}(p)=S_{\mathcal{P}_1}(p_1)\int\frac{d^4q}{(2\pi)^4}K(P,p,q)\chi_{P}(q)S_{\mathcal{P}_2}(p_2),
\end{equation}
where $S_{\mathcal{P}_1}(p_1)$ and $S_{\mathcal{P}_2}(p_2)$ are the propagators of constituent particles, and $K(P,p,q)$ is the kernel, which is defined as the sum of all the two-particle irreducible diagrams.

In the following we use the variables $p_l (=p\cdot v)$ and $p_t(=p- p_lv)$ as the longitudinal and transverse projections of the relative momentum ($p$) along the bound state velocity ($v$), respectively. Then, the propagators of the constituent mesons can be expressed as
\begin{equation}\label{propagator1}
  S_{\mathcal{P}_1}(\lambda_1P+p)=\frac{i}{\left(\lambda_1M+p_l\right)^2-\omega_1^2+i\epsilon},
\end{equation}
and
\begin{equation}\label{propagator2}
  S_{\bar{\mathcal{P}}_2}(\lambda_2P-p)=\frac{i}{\left(\lambda_2M-p_l\right)^2-\omega_2^2+i\epsilon},
\end{equation}
where $\omega_{1(2)} = \sqrt{m_{1(2)}^2+p_t^2}$ (we have defined $p_t^2=-p_t\cdot p_t$).

To obtain the interaction kernel of the two-pseudoscalar meson systems through exchanging light and heavy vector mesons, and light scalar meson, the following effective Lagrangians as in Ref. \cite{Ding:2020dio,Branz:2008ha,Li:2012as} are needed:
\begin{equation}
\begin{split}\label{Lagrangian}
   \mathcal{L}_{KK\mathbb{V}}=& ig_{KK\rho}\vec{\rho}^\mu\cdot\left(K^\dag\vec{\tau}\partial_\mu K-\partial_\mu K^\dag\vec{\tau} K\right)+i\left(g_{KK\omega}\omega^\mu+g_{KK\omega}\phi^\mu\right)(K^\dag\partial_\mu K-\partial_\mu K^\dag K),\\
   \mathcal{L}_{DD\mathbb{V}}=& ig_{{DD}\mathbb{V}}(D_b \partial_\alpha D_a^\dagger-D_a^\dagger\partial_\alpha D_b)\mathbb{V}^\alpha_{ba}+ig_{DDJ/\psi}\left(D \partial_\alpha D^\dagger-D^\dagger\partial_\alpha D\right)J/\psi^\alpha,\\
   \mathcal{L}_{BB\mathbb{V}}=& ig_{{BB}\mathbb{V}}(B_b \partial_\alpha B_a^\dagger-B_a^\dagger\partial_\alpha B_b)\mathbb{V}^\alpha_{ba}+ig_{BB\Upsilon}\left(B \partial_\alpha B^\dagger-B^\dagger\partial_\alpha B\right)\Upsilon^\alpha,\\
\mathcal{L}_{DD\sigma}=& g_{DD\sigma}D_aD^\dag_a\sigma, \,\,\,\,\,\,\mathcal{L}_{BB\sigma}= g_{BB\sigma}B_aB^\dag_a\sigma\\
  \end{split}
\end{equation}
where $J/\psi^\alpha$, $\Upsilon^\alpha$, and $\sigma$ represent the $J/\psi$, $\Upsilon$, and $\sigma$ field operators, and the nonet vector meson matrix reads as
\begin{eqnarray}
\mathbb{V}&=&\left(\begin{array}{ccc}
\frac{\rho^{0}}{\sqrt{2}}+\frac{\omega}{\sqrt{2}}&\rho^{+}&K^{*+}\\
\rho^{-}&-\frac{\rho^{0}}{\sqrt{2}}+\frac{\omega}{\sqrt{2}}&
K^{*0}\\
K^{*-} &\bar{K}^{*0}&\phi
\end{array}\right).\label{vector}
\end{eqnarray}
The coupling constants involved in Eq. (\ref{Lagrangian}) are taken as $g_{KK\rho}=g_{KK\omega}=g_{KK\phi}=3$, $g_{DD\mathbb{V}}=g_{BB\mathbb{V}}=\frac{\beta g_v}{\sqrt{2}}$ with $g_v = 5.8$, $\beta = 0.9$, $g_{DDJ/\phi}=m_{J/\psi}/f_{J/\psi}$ with $f_{J/\psi}=405$ MeV, and $g_{BB\Upsilon}=m_{\Upsilon}/f_{\Upsilon}$ with $f_{\Upsilon}=715.2$ MeV.

In the so-called ladder approximation, the interaction kernel $K(P,p,q)$ can be derived in the lowest-order form as following:
\begin{equation}\label{kernel}
\begin{split}
K(p_1,p_2;q_1,q_1,m_V)&=-(2\pi)^2\delta^4(q_1+q_2-p_1-p_2)C_Ig_{PPV}g_{P'P'V}(p_1+q_1)_\mu(p_2+q_2)_\nu\Delta^{\mu\nu}(k,m_V),\\
K(p_1,p_2;q_1,q_1,m_\sigma)&=-(2\pi)^2\delta^4(q_1+q_2-p_1-p_2)C_Ig_{PP\sigma}^2\Delta_\sigma(k,m_\sigma),
\end{split}
\end{equation}
where $m_V$ represent the masses of the exchanged light and heavy vector mesons ($\rho$, $\omega$, $\psi$, $J/\psi$, and $\Upsilon$).  $\Delta^{\mu\nu}(k,m_V)$ and $\Delta_\sigma(k,m_\sigma)$  represent the  propagators for the vector and the scalar mesons, respectively,  and they have the following forms:
\begin{equation}
\begin{split}
\Delta^{\mu\nu}(k,m_V)&=\frac{-i}{k^2-m_V^2}\left(g^{\mu\nu}-\frac{k^\mu k^\nu}{m_V^2}\right),\\
\Delta_\sigma(k,m_\sigma)&=\frac{i}{k^2-m_\sigma^2}.
\end{split}
\end{equation}
The $C_I$ in Eq. (\ref{kernel}) is the isospin coefficient for $I=0$ and $I=1$. For the $K\bar{K}$, $DK$, $\bar{B}{K}$, $D\bar{D}$, $B\bar{B}$ and $BD$ systems,
\begin{equation}\label{coe0}
    C_0=\left\{\begin{array}{rl}3/2&~~{\rm for}~\rho\\1/2&~~{\rm for}~\omega\\1&~~{\rm for}~\phi\\1&~~{\rm for}~J/\psi\\1&~~{\rm for}~\Upsilon\\1&~~{\rm for}~\sigma\end{array}\right.,~~~~~~~~~~~~~~~~~~
    C_1=\left\{\begin{array}{rl}-1/2&~~{\rm for}~\rho\\1/2&~~{\rm for}~\omega\\1&~~{\rm for}~\phi\\1&~~{\rm for}~J/\psi\\1&~~{\rm for}~\Upsilon\\1&~~{\rm for}~\sigma\end{array}\right. .
\end{equation}
For the $\bar{K}\bar{K}$, $D\bar{K}$, $\bar{B}\bar{K}$, $DD$, $\bar{B}\bar{B}$ and $\bar{B}D$ systems,
\begin{equation}\label{coe1}
C_0=\left\{\begin{array}{rl}3/2&~~{\rm for}~\rho\\-1/2&~~{\rm for}~\omega\\-1&~~{\rm for}~\phi\\-1&~~{\rm for}~J/\psi\\-1&~~{\rm for}~\Upsilon\\1&~~{\rm for}~\sigma\end{array}\right.,~~~~~~~~~~~~~~~~~~C_1=\left\{\begin{array}{rl}-1/2&~~{\rm for}~\rho\\-1/2&~~{\rm for}~\omega\\-1&~~{\rm for}~\phi\\-1&~~{\rm for}~J/\psi\\-1&~~{\rm for}~\Upsilon\\1&~~{\rm for}~\sigma\end{array}\right..
\end{equation}
In Eqs. (\ref{coe0}) and (\ref{coe1}) the exchanged mesons of $\phi$, $J/\psi$ and $\Upsilon$ only appear for the $K\bar{K}$/$\bar{K}\bar{K}$, $D\bar{D}$/$DD$ and $B\bar{B}$/$\bar{B}\bar{B}$ systems, and $\sigma$ is only considered in the doubly heavy pseudoscalar meson systems.

In order to manipulate the off shell effect of the exchanged mesons and finite size effect of the interacting hadrons, we introduce a form factor $\mathcal{F}(k^2)$ at each vertex. Generally, the form factor has the following form:
\begin{equation}\label{monopoleFF}
\mathcal{F}_M(k^2)=\frac{\Lambda^2-m^2}{\Lambda^2-k^2},
\end{equation}
where $\Lambda$, $m$ and $k$ represent the cutoff parameter, mass and momentum of the exchanged meson, respectively. This form factor is normalized at the on shell momentum of $k^2=m^2$. On the other hand, if $k^2$ were taken to be infinitely large ($-\infty$), the form factor, which can be expressed as the overlap integral of the wave functions of the hadrons at the vertex, would approach zero. Considering the difference in the wave functions and masses of the light and heavy mesons, and ensuring a positive form factor, different magnitudes of cutoff $\Lambda$ will be chosen for the heavy and light mesons.

Substituting the propagators (\ref{propagator1}) and (\ref{propagator2}), the interaction kernel (\ref{kernel}), and the form factor (\ref{monopoleFF}) into the BS Eq. (\ref{BS-equation}) and considering the instantaneous approximation ($p_l=q_l$, which mesons the energy exchanged between the constituent particles of the binding system is neglected.) in the kernel. Then, the BS Eq. (\ref{BS-equation}) with exchanging a vector meson and a scalar meson in the center-of-mass frame of the bound state ($\vec{P} = 0$) becomes
\begin{equation}\label{Three-BSeq}
\begin{split}
\chi_P(p_l,\vec{p}_t)=&\frac{i}{\left[\left(\lambda_1M+p_l\right)^2-\omega_1^2+i\epsilon\right]\left[\left(\lambda_2M-p_l\right)^2-\omega_2^2+i\epsilon\right]}\int\frac{dq_l}{2\pi}\frac{d^3\vec{q}_t}{(2\pi)^3}\\
\times&\Bigg\{C_Ig_{PPV}g_{P'P'V}\frac{4\left(\lambda_1M+p_l\right)\left(\lambda_2-p_l\right)+\left(\vec{p}_t+\vec{q}_t\right)^2+\left(\vec{p}_t^2-\vec{q}_t^2\right)/m_V^2}{-\left(\vec{p}_t-\vec{q}_t\right)-m_V^2}
\frac{\left(\Lambda^2-m_V^2\right)^2}{\left[\Lambda^2+\left(\vec{p}_t+\vec{q}_t\right)^2\right]^2}\\
&+C_Ig_{PP\sigma}g_{P'P'\sigma}\frac{1}{-\left(\vec{p}_t-\vec{q}_t\right)-m_\sigma^2}\frac{\left(\Lambda^2-m_\sigma^2\right)^2}{\left[\Lambda^2+\left(\vec{p}_t+\vec{q}_t\right)^2\right]^2}\Bigg\}\chi_P(q_l,\vec{q}_t).
\end{split}
\end{equation}
In the above equation, there are poles in the $p_l$ plane at $-\lambda_1M-\omega_1+i\epsilon$, $-\lambda_1M+\omega_1-i\epsilon$, $\lambda_2M+\omega_2-i\epsilon$ and $\lambda_2M-\omega_2+i\epsilon$. After integrating the $p_l$ on both sides of Eq. (\ref{Three-BSeq}) by selecting the proper contour, we can obtain the three-dimensional integral equation for $\tilde{\chi}_P(\vec{p}_t)$ ($\tilde{\chi}_P(\vec{p}_t)= \int dp_l\chi_P(p_l,\vec{p}_t)$), which only depends on the the three momentum, $\vec{p}_t$. By completing the azimuthal integration, the three-dimensional BS equation becomes a one-dimensional integral equation as
\begin{equation}
\tilde{\chi}_P(|\vec{p}_t|)=\int d|\vec{p}_t|A\left(|\vec{p}_t|,|\vec{q}_t|\right)\tilde{\chi}_P(|\vec{q}_t|),
\end{equation}
where the propagators and kernels after one-dimensional simplification are included in $A\left(|\vec{p}_t|,|\vec{q}_t|\right)$. The numerical solutions for $\tilde{\chi}_P(|\vec{p}_t|)$ can be obtained by discretizing the integration region into $n$ pieces (with $n$ sufficiently large). In this way, the integral equation becomes a matrix equation and the BS scalar function $\tilde{\chi}_P(|\vec{p}_t|)$ becomes $n$ dimensional vector.

\section{Numerical results}
\label{Num}
In this section, we will solve the BS equation numerically and study whether the $S$-wave bound states composed of two pseudoscalar mesons exist or not. In our model, there is only one parameter, the cutoff $\Lambda$, which comes from the form factor. The binding energy $E_b$ is defined as $E_b=M-m_1-m_2$ in the rest frame of the bound state. We take the averaged masses of the pseudoscalar mesons and the exchanged light and heavy mesons from PDG \cite{pdg2020}, $m_K$ = 494.988 MeV, $m_D$ = 1868.04 MeV, $m_{B}$ = 5279.44 MeV, $m_{\rho}$ = 775.26 MeV, $m_{\omega}$ = 782.65 MeV, $m_{\phi}$ = 1019.461 MeV, $m_{J/\psi}$ = 3096.9 MeV, and $m_\Upsilon$ = 9460.3 MeV.

\begin{figure}[htbp]
\centering
\subfigure[]{
\begin{minipage}[t]{0.5\linewidth}
\centering
\includegraphics[width=3.2in]{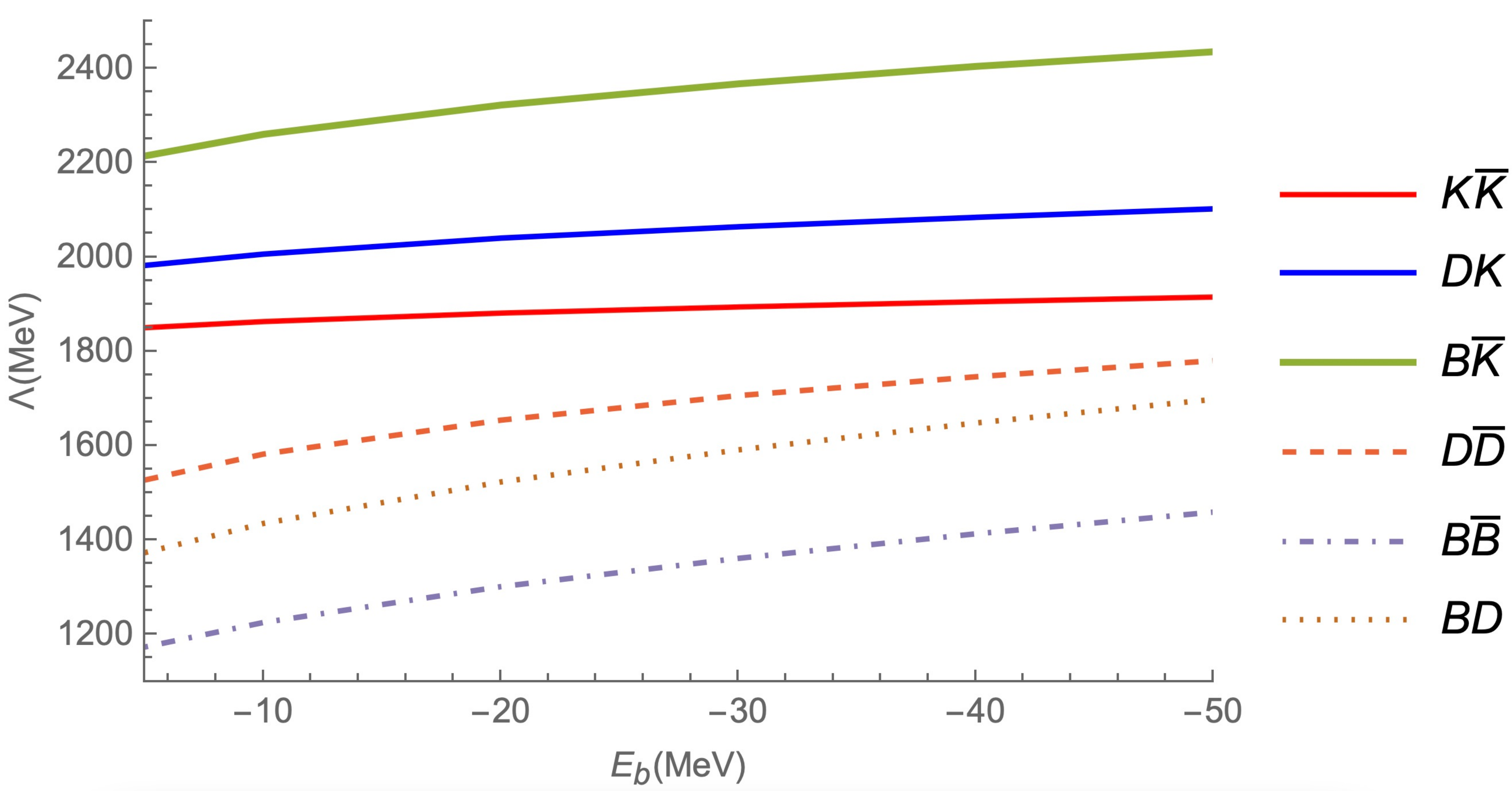}
\end{minipage}%
}%
\subfigure[]{
\begin{minipage}[t]{0.5\linewidth}
\centering
\includegraphics[width=3in]{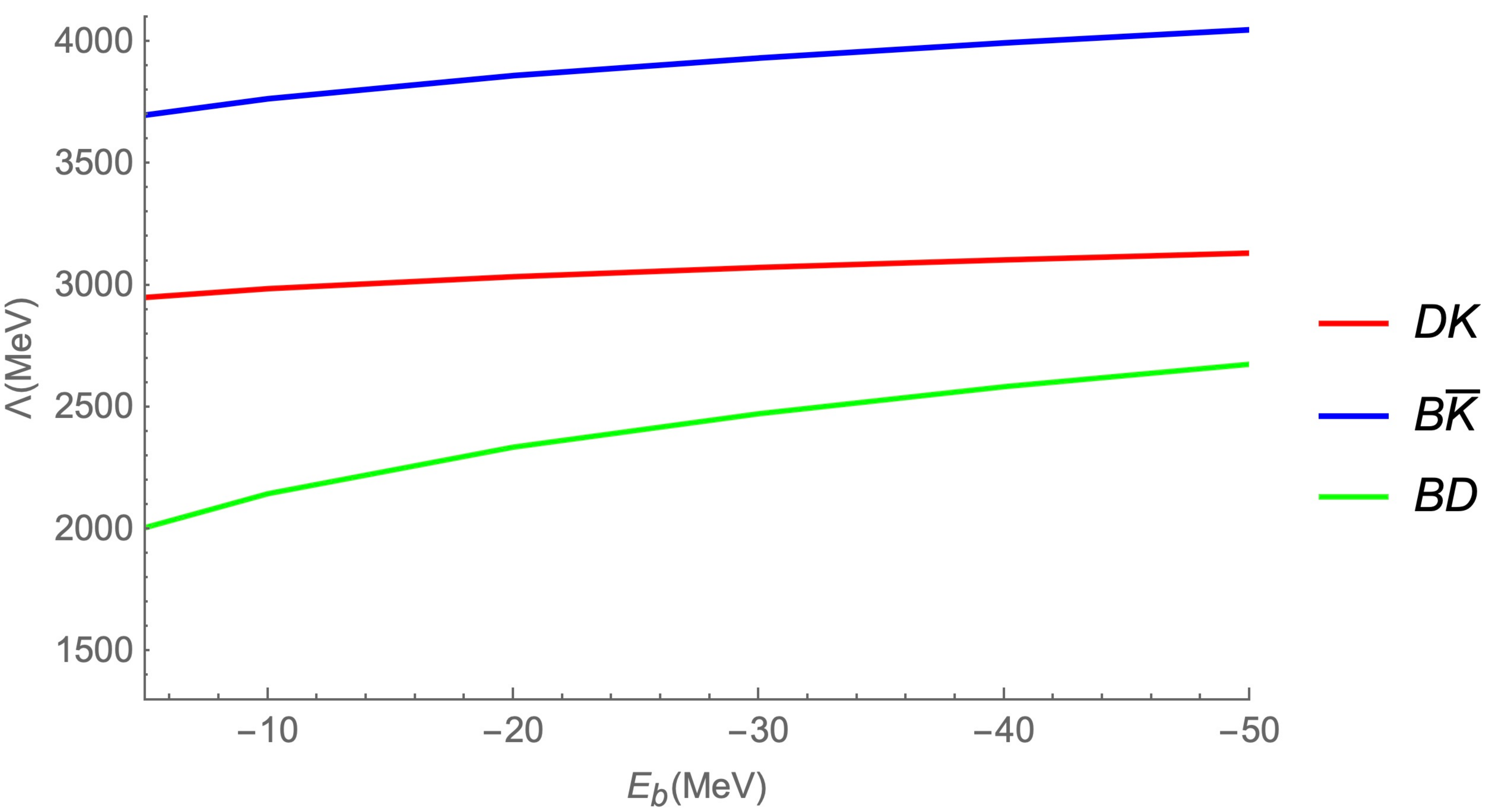}
\end{minipage}
}%
\centering
\caption{The numerical results for $K\bar{K}$, $DK$, $\bar{B}{K}$, $D\bar{D}$, $B\bar{B}$, and $BD$ systems with $I=0$ (a) and $D\bar{K}$, $\bar{B}\bar{K}$, and $\bar{B}D$ systems with $I=0$ (b).}
\label{NR}
\end{figure}

In Fig. \ref{NR}, we present some possible bound states composed of two pseudoscalar mesons when only the light meson ($\rho$, $\omega$, $\phi$, and $\sigma$) exchange contributions are considered. Here we vary the binding energy from 0 to -50 MeV and the cutoff in a wide range (0.8-5) GeV. We find that only the $K\bar{K}$, $DK$, $\bar{B}{K}$, $D\bar{D}$, $B\bar{B}$, $BD$, $D\bar{K}$, $\bar{B}\bar{K}$, $\bar{B}D$ systems with $I=0$ can exist as bound states. For the $\bar{K}\bar{K}$, $DD$, and $\bar{B}\bar{B}$ systems with $I=0$ are forbidden because of the Bose symmetry and the interactions in $I=1$ systems are repulsive, hence no bound states exist in the $\bar{K}\bar{K}$, $DD$, and $\bar{B}\bar{B}$ systems. Furthermore, we cannot predict with certainty masses of bound states which will be measured experimentally due to that our results are dependent on the cutoff $\Lambda$. The contribution of the $\sigma$ exchange is included in our work, despite the large uncertainties in its mass and structure. In our previous works \cite{Zhao:2021cvg,Wang:2020lua} and Ref. \cite{Ding:2008gr} it was found that the contribution of $\sigma$ exchange is very small to form bound states, and the same result is found in our current work.

The systems that may exist as bound states are presented in Fig. \ref{NR} . It is noted that in the hidden bottom system the cutoff is the smallest in Fig. \ref{NR}(a). This is because the mass of $B$ meson is the largest one, which requires smaller cutoff value as compared to the other systems, and the cutoff is determined by the overlap integrals of the wave functions the hadrons at the vertices and the size of $B$ meson is the smallest among the constituent particles. There is no bound state for the system with $I=1$. This is because the isospin coefficients of $\rho$ and $\omega$ are -1/2 and 1/2, respectively, as shown in Eq. (\ref{coe0}) and the masses of $\rho$ and $\omega$ are almost equal, leading to the contributions from $\rho$ and $\omega$ exchanges almost canceling each other. Among these possible bound states, the $K\bar{K}$ and $DK$ bound states can be related to the experimentally observed $f_0(980)$ and $D_{s0}^\ast(2317)$, respectively \cite{Guo:2017jvc}. Based on the heavy chiral unitary approach \cite{Guo:2006fu} and the linear chiral symmetry \cite{Bardeen:2003kt}, the authors predicted the existence of a $b$-partner state $B_{s0}^\ast$ of $D_{s0}^\ast(2317)$ as the $B\bar{K}$ bound state, which also can be confirmed in our model with the cutoff $\Lambda$ = 2436 MeV. For the experimentally observed $X(5568)$, it cannot be a $B\bar{K}$ bound state in our model \cite{Wang:2018jsr}. In Ref. \cite{Gamermann:2006nm}, a new hidden charm resonance with mass 3.7 GeV (named as $X(3700)$) was predicted corresponding mostly to a $D\bar{D}$ state. Later it has been searched in $B\rightarrow D\bar{D}K$ \cite{Dai:2015bcc}, $e^+e^-\rightarrow J/\psi D\bar{D}$ \cite{Xiao:2012iq}, $\psi(3770)\rightarrow X(3700)\gamma$ \cite{Gamermann:2009ouq}, $\gamma\gamma\rightarrow D\bar{D}$ \cite{Wang:2020elp,Deineka:2021aeu}, $\Lambda_b\rightarrow\Lambda D\bar{D}$ \cite{Wei:2021usz}, etc. Recently, lattice QCD also found a $D\bar{D}$ bound state just below the threshold with the binding energy $E_b=-4.0^{+3.7}_{-5.0}$ MeV \cite{Prelovsek:2020eiw}. The existence of the $B\bar{B}$ bound state was also confirmed by the effective potential model \cite{Liu:2009qhy}, the heavy quark effective theory \cite{Liu:2017mrh}, the chiral SU(3) QM \cite{Liu:2008mi}  and the qBSE \cite{Ding:2020dio}. The $BD$ system, in analogy to the $DK$ system, can also be a bound state in the local hidden gauge symmetry (HGS) approach \cite{Sakai:2017avl}.

Not long ago, the $T_{cc}$ with the quantum numbers $I(J^P)=0(1^+)$ and the quark content $cc\bar{u}\bar{d}$ was reported by the LHCb Collaboration \cite{LHCb:2021auc}, which is the first experimentally discovered open charmed tetraquark state. The mass of $T_{cc}$ is just below the threshold of $DD^\ast$, and could be an ideal candidate for the $DD^\ast$ bound state. In fact, this inspired us to investigate the possibility of two-pseudoscalar meson systems as bound states with open flavour. In Ref. \cite{Li:2012ss}, the authors  systematically investigated possible deuteron-like molecular states with two heavy quarks by one-boson-exchange (OBE) model. According to their results, the $I=1$ $DD$ system might not be a molecule, the $I=1$ $\bar{B}\bar{B}$, $I=0$ and $I=1$ $D\bar{B}$ systems might be molecule candidates, but the results depend a little sensitively on the cutoff. Based on the Heavy-Meson Effective Theory, the $DD$ with $I=1$, $\bar{B}\bar{B}$ with $I=1$, and $D\bar{B}$ with $I=0$ and $I=1$ systems could exist as shallow bound states \cite{Abreu:2022sra}.

For the results of two-pseudoscalar meson systems only the isoscalar system can exist as bound states, which are presented in Fig. \ref{NR}(b). This is because for the isovector systems the isospin coefficients corresponding to $\rho$ and $\omega$ exchanges are  -1/2, as shown in Eq. (\ref{coe1}), so the total interaction is repulsive in the isovector systems. Comparing the results in Fig. \ref{NR} (a) and Fig. \ref{NR}(b), we can obviously see that the cutoff $\Lambda$ is larger in Fig. \ref{NR}(b), which is caused by the difference in the isospin coefficients, i.e. 1/2  and -1/2 in the systems with $I=0$ due to the $\omega$ exchange in Eq. (\ref{coe0}) and Eq. (\ref{coe1}), respectively. From Fig. \ref{NR}(a) and Fig. \ref{NR}(b), we can also find that for the constituent particles with same masses, the larger the mass of the constituent particle the smaller the cutoff $\Lambda$. And the larger difference in masses of the constituent particles the larger cutoff $\Lambda$. In order to facilitate the comparison the results of different theoretical models, we have listed the results of some different models and our results in Table \ref{DNR}.

\renewcommand{\arraystretch}{1.0}
\begin{table}[htp]
\centering \caption{The results for different theoretical models. ``$\surd$" and ``$\times$" denote that the corresponding system might be a bound state or not, respectively. ``$-$" means corresponding system without study.}
\label{DNR}
\begin{tabular*}{18cm}{@{\extracolsep{\fill}}lcccccccccccccccccccccccc}
\toprule[1.0pt]\addlinespace[3pt]
\midrule[1pt]
         & \multicolumn{2}{c}{$K\bar{K}$} & \multicolumn{2}{c}{$DK$} &\multicolumn{2}{c}{$\bar{B}K$} &\multicolumn{2}{c}{$D\bar{D}$}& \multicolumn{2}{c}{$B\bar{B}$} & \multicolumn{2}{c}{$BD$} & \multicolumn{2}{c}{$\bar{K}\bar{K}$} & \multicolumn{2}{c}{$D\bar{K}$} & \multicolumn{2}{c}{$\bar{B}\bar{K}$} &\multicolumn{2}{c}{$DD$} &\multicolumn{2}{c}{$\bar{B}\bar{B}$}  & \multicolumn{2}{c}{$\bar{B}D$} \\
\midrule[1pt]
                                                        $I$     & 0    & 1    &0     &1  & 0 & 1 & 0 & 1 &0 &1  & 0 & 1& 0 & 1 &0 &1 & 0 & 1 & 0 & 1 &0 &1  & 0 & 1\\
qBSE \cite{Kong:2021ohg,Ding:2020dio} & $-$ & $-$ &$\surd$&$\times$ &$\surd$&$\times$&$\surd$&$\times$&$\surd$& $\times$& $-$ &$-$ 
                                                                 & $-$ & $-$ &$\surd$&$\times$ &$\surd$&$\times$&$\surd$&$\times$&$\surd$& $\times$& $-$ &$-$\\                                                                      
HGS \cite{Sakai:2017avl,Dai:2022ulk} & $-$ & $-$ &$-$&$-$&$-$&$-$&$-$&$-$&$-$& $-$& $\surd$&$\times$ 
                                                                 & $-$ & $-$ &$-$&$-$ &$-$&$-$&$-$&$-$&$\times$& $\times$& $\surd$&$\times$\\ 
OBE \cite{Liu:2009qhy,Li:2012ss} & $-$ & $-$   &$-$&$-$    &$-$&$-$   &$\times$&$\surd$    &$\times$& $\surd$    & $\surd$&$\times$ 
                                                                 & $-$ & $-$    &$-$&$-$    &$-$&$-$   &$\times$&$\times$    &$\times$& $\surd$    & $\surd$&$\surd$\\ 
CCS \cite{Ohkoda:2012hv,Ohkoda:2011vj} & $-$ & $-$   &$-$&$-$    &$-$&$-$   &$-$&$-$    &$\times$& $\surd$    & $-$&$-$ 
                                                                 & $-$ & $-$    &$-$&$-$    &$-$&$-$   &$\times$&$\times$    &$\times$& $\surd$    & $-$&$-$\\   
QM \cite{Liu:2008mi} & $-$ & $-$   &$-$&$-$    &$-$&$-$   &$\surd$&$\times$    &$\surd$&$\times$    &$\surd$&$\times$
                                                                 & $-$ & $-$    &$-$&$-$    &$-$&$-$   &$-$&$-$    &$-$& $-$    & $-$&$-$\\   
NRS \cite{Zhang:2006ix} &$\times$ &$\times$   &$\surd$&$\times$   &$\surd$&$\times$   &$\surd$&$\times$    &$\surd$&$\times$    &$\surd$&$\times$
                                                                 &$\times$&$\times$    &$\surd$&$\times$   &$\times$&$\times$  &$\times$&$\times$    &$\times$&$\times$    &$\surd$&$\times$\\  
Our results &$\surd$ &$\times$   &$\surd$&$\times$   &$\surd$&$\times$   &$\surd$&$\times$    &$\surd$&$\times$    &$\surd$&$\times$
                                                                 &$\times$&$\times$    &$\surd$&$\times$   &$\surd$&$\times$  &$\times$&$\times$    &$\times$&$\times$    &$\surd$&$\times$\\                                                                                                                              
\midrule[1pt]
\bottomrule[1.0pt]
\end{tabular*}
\end{table}

In Ref. \cite{Aceti:2014uea}, the authors systematically studied the interaction of $D\bar{D}^\ast$ in the isospin $I=0$ channel. In their work it is shown the exchange of a light $q\bar{q}$ is OZI forbidden in the $I=1$ channel. As a consequence, only the $J/\psi$ exchange is allowed in the case of $I=1$, and the simultaneous two pion exchange, which was evaluated in \cite{Aceti:2014uea} and \cite{Aceti:2014kja}, was found to be weaker than the exchange of the vector meson. In spite of the large mass of the $J/\psi$, which suppresses the propagator of the exchanged $J/\psi$ exchange, it was found in \cite{Aceti:2014uea} and \cite{Aceti:2014kja} that the interaction could bind the $D\bar{D}^\ast$ and $D^\ast\bar{D}^\ast$ systems with $I=1$ weakly. Subsequently, in Ref. \cite{He:2015mja} it was also found the bound state $D\bar{D}^\ast$ with $I^G(J^P)=1^+(1^+)$ would disappear if the $J/\psi$ exchange were removed, which means that the $J/\psi$ exchange is important to provide an attractive interaction to produce the pole in the isovector system. In the present work, we also consider the effects of exchanged heavy mesons. Considering the differences in the wave functions and masses of the light and heavy mesons, and ensuring a positive from factor, we choose different magnitudes of cutoffs $\Lambda_L$ and $\Lambda_H$ for the exchanged light and heavy mesons, respectively.

\begin{figure}[htbp]
\centering
\subfigure[]{
\begin{minipage}[t]{0.5\linewidth}
\centering
\includegraphics[width=3.3in]{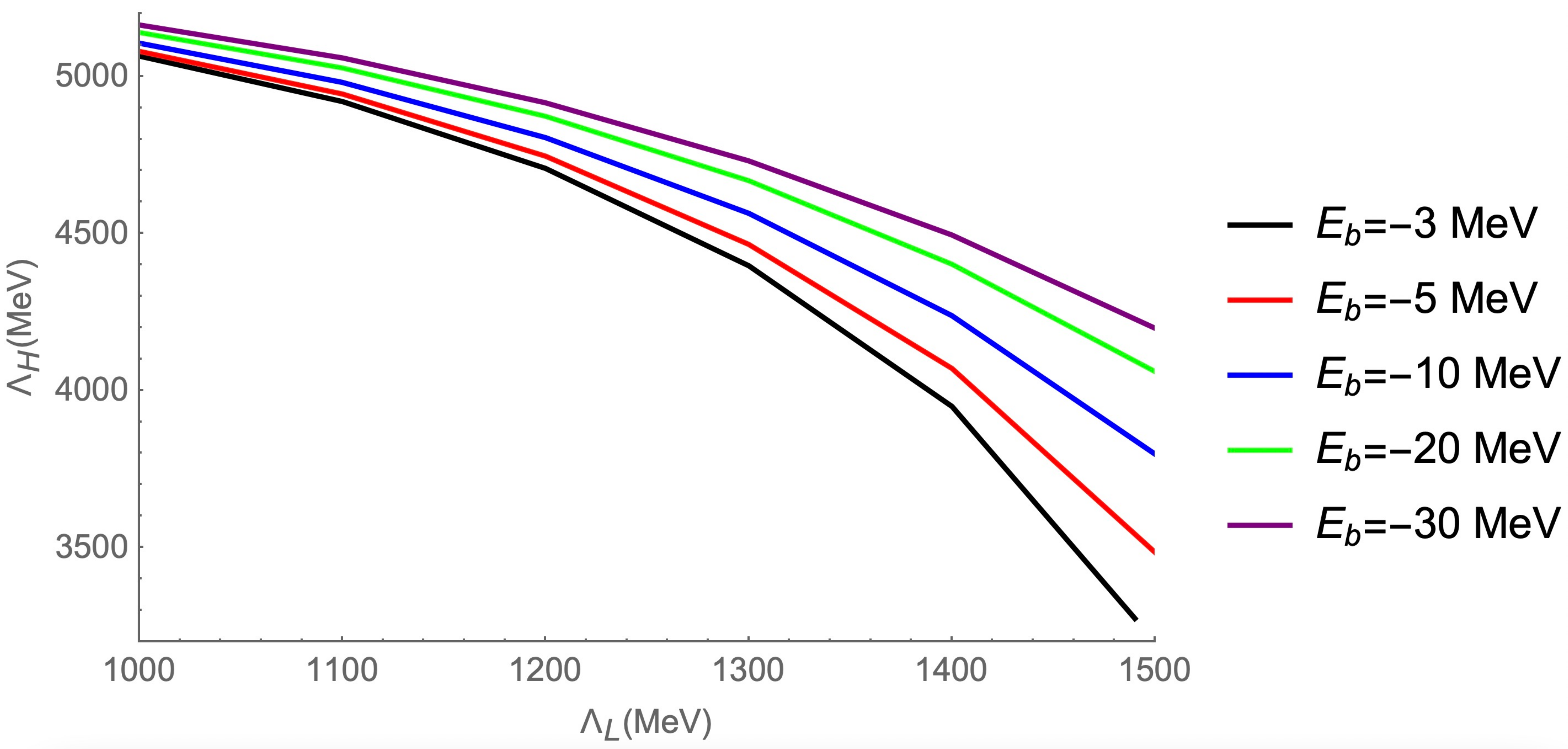}
\end{minipage}%
}%
\subfigure[]{
\begin{minipage}[t]{0.5\linewidth}
\centering
\includegraphics[width=3.3in]{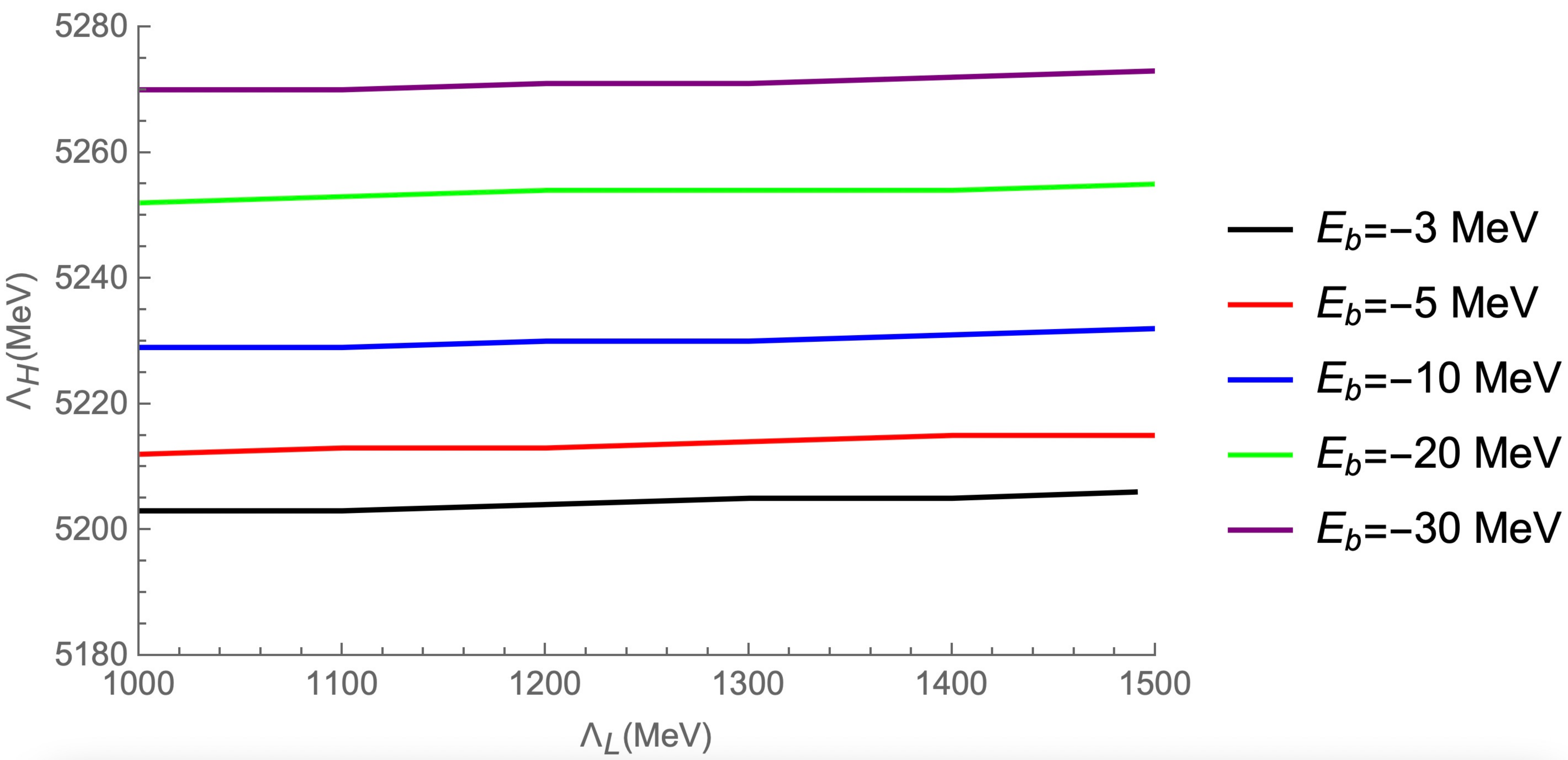}
\end{minipage}
}%
\centering
\caption{The numerical results for $I=0$ (a) and $I=1$ (b) $D\bar{D}$ systems with $J/\psi$ meson exchange included.}
\label{DD}
\end{figure}

\begin{figure}[htbp]
\centering
\subfigure[]{
\begin{minipage}[t]{0.5\linewidth}
\centering
\includegraphics[width=3.3in]{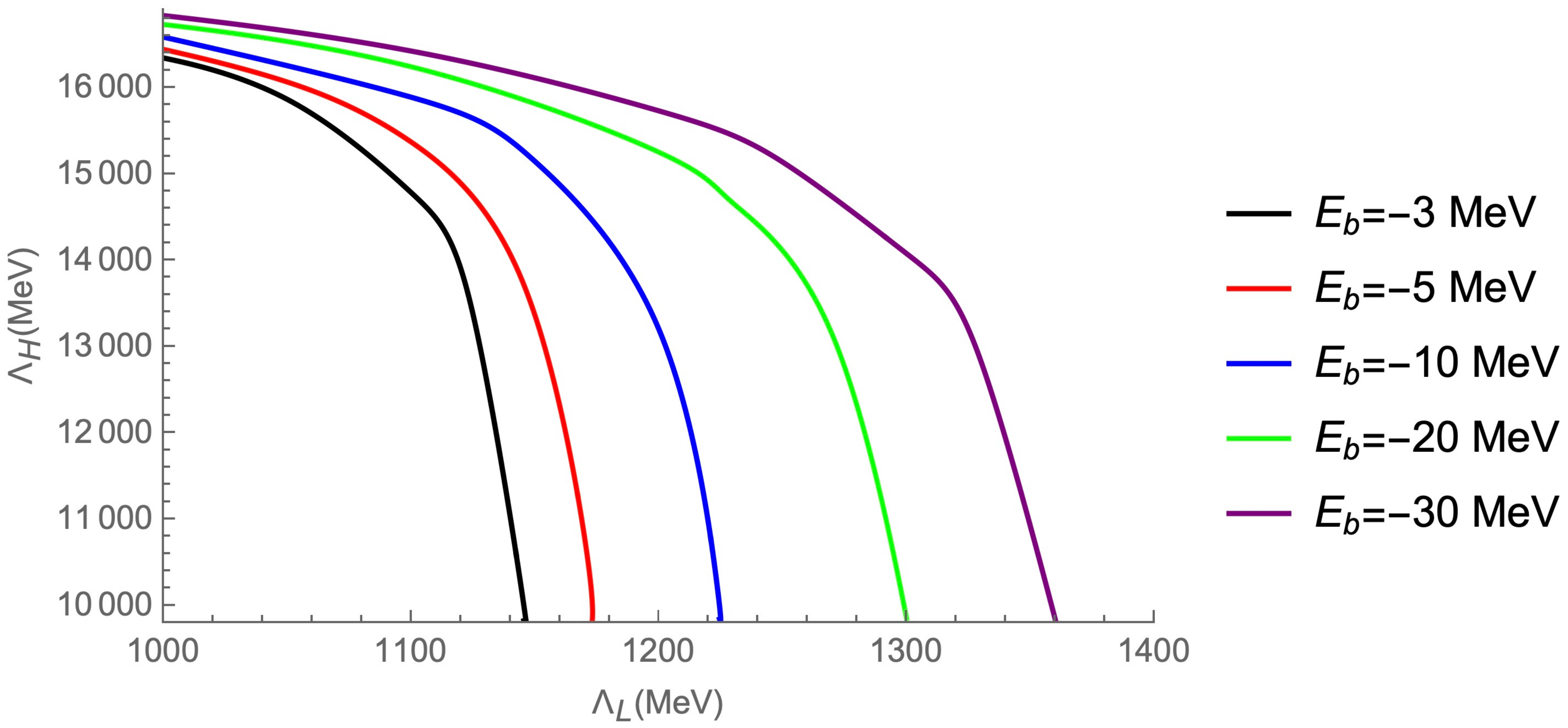}
\end{minipage}%
}%
\subfigure[]{
\begin{minipage}[t]{0.5\linewidth}
\centering
\includegraphics[width=3.3in]{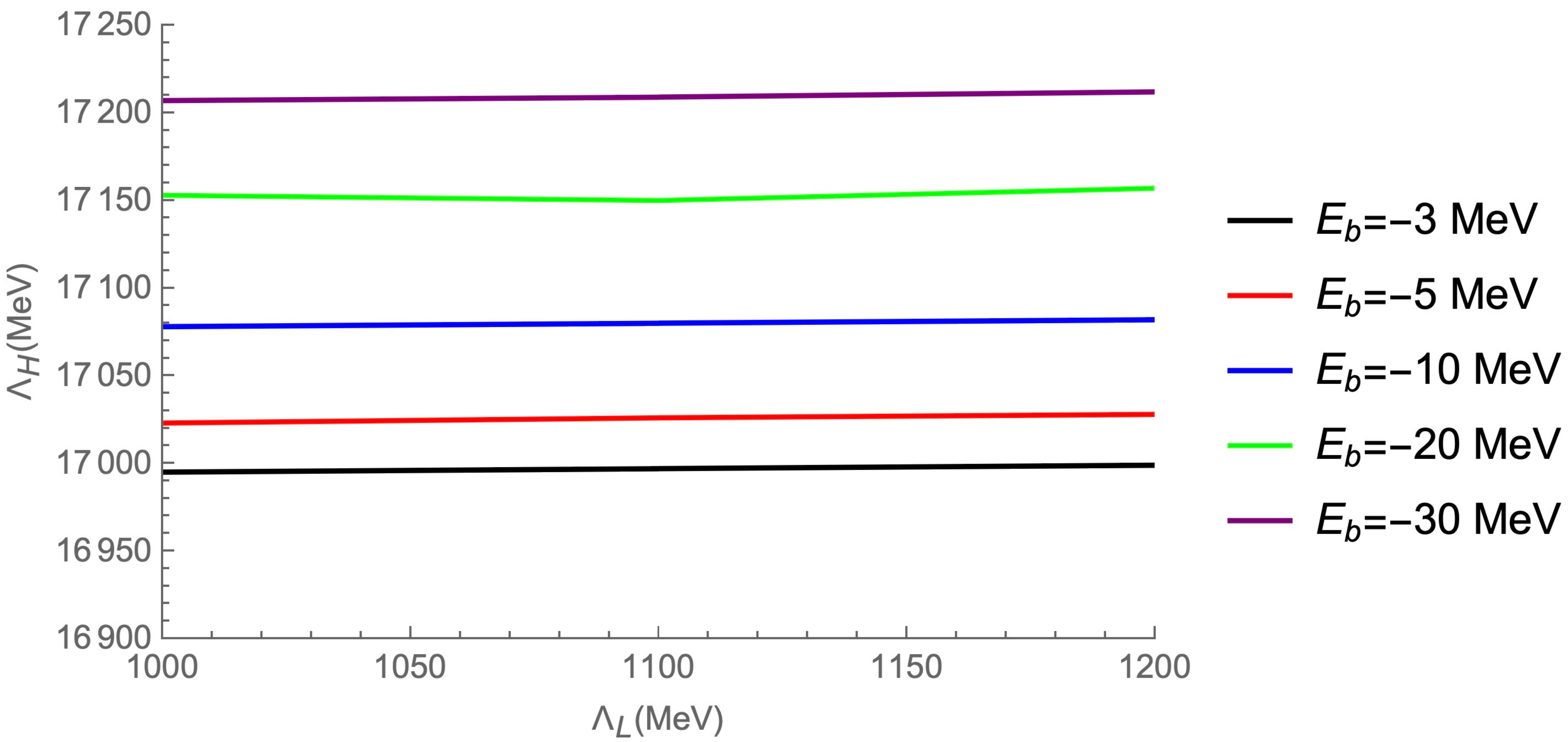}
\end{minipage}
}%
\centering
\caption{The numerical results for $I=0$ (a) and $I=1$ (b) $B\bar{B}$ systems with $\Upsilon$ meson exchange included.}
\label{BB}
\end{figure}

We vary the cutoff $\Lambda_L$ due to the exchanged light mesons in the range of (800-1500) MeV to find the cutoff $\Lambda_H$ due to the exchanged heavy mesons that can form bound states. The results for some possible bound states of $D\bar{D}$ and $B\bar{B}$ are presented in Figs. \ref{DD} and \ref{BB}. From these results in Figs \ref{DD} and \ref{BB}, we can see that the effect of the exchange of a heavy meson can not be ignored. Due to the contributions of the exchanged $\rho$ and $\omega$ almost cancel each other in the $I=1$ $D\bar{D}$ and $B\bar{B}$ systems, the main contribution comes from the heavy meson exchange. It can be seen that the results for $\Lambda_H$ are almost twice the mass of the exchanged heavy meson in Figs. \ref{DD}(b) and \ref{BB}(b). This has the same situation with only considered the light mesons exchange, the value of the cutoff $\Lambda$ are also about twice the mass of the exchange meson as in Fig. \ref{NR}(a). Whether the bound states can be formed only considered the contribution of the heavy meson exchange, the existence of $\bar{D}D$ and $B\bar{B}$ bound states with $I=1$ experimentally is determinant.

\section{summary}
In this paper we derived the BS equation for the $S$-wave $K\bar{K}$, $DK$, ${B}\bar{K}$, $D\bar{D}$, $B\bar{B}$, $BD$, $KK$, $D\bar{K}$, $\bar{B}\bar{K}$, $DD$, $\bar{B}\bar{B}$ and $\bar{B}D$ systems, and systematically studied the possible bound states of these systems with the ladder approximation and the instantaneous approximation for the kernel. In our model, the kernel containing one-particle-exchange diagrams induced by the light meson ($\rho$, $\omega$, $\phi$, and $\sigma$) and the heavy meson ($J/\psi$ and $\Upsilon$) exchanges. To investigate the bound states, we have numerically solved the BS equations for $S$-wave systems composed of two pseudoscalar mesons. The possible $S$-wave bound states studied in our work are helpful in explaining the structures of experimentally discovered exotic states and predicting unobserved exotic states.

As results, we found the $K\bar{K}$, $DK$, $\bar{B}{K}$, $D\bar{D}$, $B\bar{B}$, $BD$, $D\bar{K}$, $\bar{B}\bar{K}$, and $\bar{B}D$ with $I=0$ can exist as bound states. For the $\bar{K}\bar{K}$, $DD$, and $\bar{B}\bar{B}$ systems with $I=0$ are forbidden because of the Bose symmetry and the interactions in $I=1$ systems are repulsive, hence no bound states exist in the $\bar{K}\bar{K}$, $DD$, and $\bar{B}\bar{B}$ systems. We also found that for the constituent particles with same masses, the larger mass of the constituent particle the smaller the cutoff $\Lambda$. And the larger difference in masses of the constituent particles the larger cutoff $\Lambda$. The contribution of $\sigma$ exchange is very small to form bound states.

In the calculation, we considered the heavy meson exchanges in the kernel. We found the effect of the heavy mesons exchange can not be neglected for the $D\bar{D}$ and $B\bar{B}$ systems. Since the contributions from the $\rho$ and $\omega$ exchanges almost cancel each other in the $I=1$ $D\bar{D}$ and $B\bar{B}$ systems, the main contribution comes from the heavy meson exchanges, and the $I=1$ $D\bar{D}$ and $B\bar{B}$ systems can exist as bound states . However, since the cutoff $\Lambda_H$ for the heavy meson exchanges is very big. Whether the bound states can be formed only considered the contribution of the heavy meson exchange, the existence of $\bar{D}D$ and $B\bar{B}$ bound states with $I=1$ experimentally is determinant.

With the restarted LHC and other experiments, more experimental studies of exotic hadrons will be performed in the near future. Recently, the LHCb collaboration observed three never-before-seen particles: a new kind of pentaquark and the first-ever pair of tetraquarks, which includes a new type of tetraquark \cite{LHCb}. These will help physicists better understand how quarks bind together into exotic particles. There is still controversy about the theoretical explanation of the structures of experimentally observed exotic hadrons and the existence of possible molecular states predicted theoretically. Therefore, more precise experimental studies of the exotic states will be needed to test the results of theoretical studies and to improve theoretical models.

\acknowledgments
One of the authors (Z.-Y. Wang) thanks Professor Jia-Jun Wu and Dr. Rui-Cheng Li for helpful discussions and useful suggestions. This work was supported by National Natural Science Foundation of China (Projects No. 12105149 and No. 11775024).


\begin{thebibliography}{99}
\bibitem{Han:1965pf}
M.~Y.~Han and Y.~Nambu,
Phys. Rev. \textbf{139}, B1006-B1010 (1965).

\bibitem{ParticleDataGroup:2020ssz}
P.~A.~Zyla \textit{et al.} [Particle Data Group],
PTEP \textbf{2020}, 083C01 (2020).



\bibitem{Astier:1967zz}
A.~Astier, L.~Montanet, M.~Baubillier and J.~Duboc,
Phys. Lett. B \textbf{25}, 294-297 (1967).

\bibitem{Ammar:1968zur}
R.~Ammar, R.~Davis, W.~Kropac, J.~Mott, D.~Slate, B.~Werner, M.~Derrick, T.~Fields and F.~Schweingruber,
Phys. Rev. Lett. \textbf{21}, 1832-1835 (1968).

\bibitem{BaBar:2003oey}
B.~Aubert \textit{et al.} [BaBar],
Phys. Rev. Lett. \textbf{90}, 242001 (2003).

\bibitem{Godfrey:2015dva}
S.~Godfrey and K.~Moats,
Phys. Rev. D \textbf{93}, 034035 (2016).

\bibitem{Olsen:2017bmm}
S.~L.~Olsen, T.~Skwarnicki and D.~Zieminska,
Rev. Mod. Phys. \textbf{90}, 015003 (2018).



\bibitem{Weinstein:1982gc}
J.~D.~Weinstein and N.~Isgur,
Phys. Rev. Lett. \textbf{48}, 659 (1982).

\bibitem{Weinstein:1983gd}
J.~D.~Weinstein and N.~Isgur,
Phys. Rev. D \textbf{27}, 588 (1983).

\bibitem{Weinstein:1990gu}
J.~D.~Weinstein and N.~Isgur,
Phys. Rev. D \textbf{41}, 2236 (1990).

\bibitem{Oller:2002na}
J.~A.~Oller,
Nucl. Phys. A \textbf{714}, 161-182 (2003).

\bibitem{Baru:2003qq}
V.~Baru, J.~Haidenbauer, C.~Hanhart, Y.~Kalashnikova and A.~E.~Kudryavtsev,
Phys. Lett. B \textbf{586}, 53-61 (2004).

\bibitem{Dai:2014zta}
L.~Y.~Dai and M.~R.~Pennington,
Phys. Rev. D \textbf{90}, 036004 (2014).

\bibitem{Ahmed:2020kmp}
H.~A.~Ahmed and C.~W.~Xiao,
Phys. Rev. D \textbf{101}, 094034 (2020).

\bibitem{Wang:2022vga}
Z.~Q.~Wang, X.~W.~Kang, J.~A.~Oller and L.~Zhang,
Phys. Rev. D \textbf{105}, 074016 (2022).



\bibitem{Szczepaniak:2003vy}
A.~P.~Szczepaniak,
Phys. Lett. B \textbf{567}, 23-26 (2003).

\bibitem{Hofmann:2003je}
J.~Hofmann and M.~F.~M.~Lutz,
Nucl. Phys. A \textbf{733}, 142-152 (2004).

\bibitem{Guo:2006fu}
F.~K.~Guo, P.~N.~Shen, H.~C.~Chiang, R.~G.~Ping and B.~S.~Zou,
Phys. Lett. B \textbf{641}, 278-285 (2006).

\bibitem{Flynn:2007ki}
J.~M.~Flynn and J.~Nieves,
Phys. Rev. D \textbf{75}, 074024 (2007).

\bibitem{Xie:2010zza}
Z.~X.~Xie, G.~Q.~Feng and X.~H.~Guo,
Phys. Rev. D \textbf{81}, 036014 (2010).

\bibitem{Guo:2015dha}
Z.~H.~Guo, U.~G.~Mei\ss{}ner and D.~L.~Yao,
Phys. Rev. D \textbf{92}, 094008 (2015).

\bibitem{Du:2017ttu}
M.~L.~Du, F.~K.~Guo, U.~G.~Mei\ss{}ner and D.~L.~Yao,
Eur. Phys. J. C \textbf{77}, 728 (2017).

\bibitem{Wu:2019vsy}
T.~W.~Wu, M.~Z.~Liu, L.~S.~Geng, E.~Hiyama and M.~P.~Valderrama,
Phys. Rev. D \textbf{100}, 034029 (2019).

\bibitem{Kong:2021ohg}
S.~Y.~Kong, J.~T.~Zhu, D.~Song and J.~He,
Phys. Rev. D \textbf{104}, 094012 (2021).

\bibitem{Huang:2021fdt}
B.~L.~Huang, Z.~Y.~Lin and S.~L.~Zhu,
Phys. Rev. D \textbf{105}, 036016 (2022).

\bibitem{Zhang:2006ix}
Y.~J.~Zhang, H.~C.~Chiang, P.~N.~Shen and B.~S.~Zou,
Phys. Rev. D \textbf{74}, 014013 (2006).

\bibitem{Faessler:2008vc}
A.~Faessler, T.~Gutsche, V.~E.~Lyubovitskij and Y.~L.~Ma,
Phys. Rev. D \textbf{77}, 114013 (2008).

\bibitem{Feng:2011zzb}
G.~Q.~Feng, Z.~X.~Xie and X.~H.~Guo,
Phys. Rev. D \textbf{83}, 016003 (2011).

\bibitem{D0:2016mwd}
V.~M.~Abazov \textit{et al.} [D0],
Phys. Rev. Lett. \textbf{117}, 022003 (2016).

\bibitem{Xiao:2016mho}
C.~J.~Xiao and D.~Y.~Chen,
Eur. Phys. J. A \textbf{53}, 127 (2017).

\bibitem{Agaev:2016urs}
S.~S.~Agaev, K.~Azizi and H.~Sundu,
Eur. Phys. J. Plus \textbf{131}, 351 (2016).

\bibitem{Chen:2016npt}
X.~Chen and J.~Ping,
Eur. Phys. J. C \textbf{76}, 351 (2016).

\bibitem{Chen:2016ypj}
R.~Chen and X.~Liu,
Phys. Rev. D \textbf{94}, 034006 (2016).

\bibitem{Lu:2016kxm}
J.~X.~Lu, X.~L.~Ren and L.~S.~Geng,
Eur. Phys. J. C \textbf{77}, 94 (2017).

\bibitem{Wang:2018jsr}
Z.~Y.~Wang, J.~J.~Qi and X.~H.~Guo,
Adv. High Energy Phys. \textbf{2019}, 7576254 (2019).

\bibitem{Liu:2008mi}
Y.~R.~Liu and Z.~Y.~Zhang,
Phys. Rev. C \textbf{80}, 015208 (2009).

\bibitem{Ohkoda:2012hv}
S.~Ohkoda, Y.~Yamaguchi, S.~Yasui, K.~Sudoh and A.~Hosaka,
Phys. Rev. D \textbf{86}, 034019 (2012).

\bibitem{Ding:2020dio}
Z.~M.~Ding, H.~Y.~Jiang and J.~He,
Eur. Phys. J. C \textbf{80}, 1179 (2020).

\bibitem{Gamermann:2006nm}
D.~Gamermann, E.~Oset, D.~Strottman and M.~J.~Vicente Vacas,
Phys. Rev. D \textbf{76}, 074016 (2007).

\bibitem{Dai:2015bcc}
L.~R.~Dai, J.~J.~Xie and E.~Oset,
Eur. Phys. J. C \textbf{76}, 121 (2016).

\bibitem{Dai:2020yfu}
L.~Dai, G.~Toledo and E.~Oset,
Eur. Phys. J. C \textbf{80}, 510 (2020).

\bibitem{Wang:2020elp}
E.~Wang, H.~S.~Li, W.~H.~Liang and E.~Oset,
Phys. Rev. D \textbf{103}, 054008 (2021).

\bibitem{Deineka:2021aeu}
O.~Deineka, I.~Danilkin and M.~Vanderhaeghen,
Phys. Lett. B \textbf{827}, 136982 (2022).

\bibitem{Aceti:2014kja}
F.~Aceti, M.~Bayar, J.~M.~Dias and E.~Oset,
Eur. Phys. J. A \textbf{50}, 103 (2014).

\bibitem{Aceti:2014uea}
F.~Aceti, M.~Bayar, E.~Oset, A.~Martinez Torres, K.~P.~Khemchandani, J.~M.~Dias, F.~S.~Navarra and M.~Nielsen,
Phys. Rev. D \textbf{90}, 016003 (2014).

\bibitem{Ding:2021igr}
Z.~M.~Ding, H.~Y.~Jiang, D.~Song and J.~He,
Eur. Phys. J. C \textbf{81}, 732 (2021).

\bibitem{Branz:2008ha}
T.~Branz, T.~Gutsche and V.~E.~Lyubovitskij,
Phys. Rev. D \textbf{78}, 114004 (2008).

\bibitem{Li:2012as}
G.~Li, F.~l.~Shao, C.~W.~Zhao and Q.~Zhao,
Phys. Rev. D \textbf{87}, 034020 (2013).

\bibitem{pdg2020}
Particle Data Group, P.~Zyla et al., Review of particle physics, Prog. Theor. Exp.
Phys. \textbf{6}, 083C01 (2020).

\bibitem{Zhao:2021cvg}
M.~J.~Zhao, Z.~Y.~Wang, C.~Wang and X.~H.~Guo,
Phys. Rev. D \textbf{105}, 096016 (2022).

\bibitem{Wang:2020lua}
Z.~Y.~Wang, J.~J.~Qi, J.~Xu and X.~H.~Guo,
Phys. Rev. D \textbf{102}, 036008 (2020).

\bibitem{Ding:2008gr}
G.~J.~Ding,
Phys. Rev. D \textbf{79}, 014001 (2009).


\bibitem{Guo:2017jvc}
F.~K.~Guo, C.~Hanhart, U.~G.~Mei\ss{}ner, Q.~Wang, Q.~Zhao and B.~S.~Zou,
Rev. Mod. Phys. \textbf{90}, 015004 (2018).

\bibitem{Bardeen:2003kt}
W.~A.~Bardeen, E.~J.~Eichten and C.~T.~Hill,
Phys. Rev. D \textbf{68} (2003), 054024.

\bibitem{Dai:2015bcc}
L.~R.~Dai, J.~J.~Xie and E.~Oset,
Eur. Phys. J. C \textbf{76}, 121 (2016).

\bibitem{Xiao:2012iq}
C.~W.~Xiao and E.~Oset,
Eur. Phys. J. A \textbf{49}, 52 (2013).

\bibitem{Gamermann:2009ouq}
D.~Gamermann, E.~Oset and B.~S.~Zou,
Eur. Phys. J. A \textbf{41}, 85-91 (2009).

\bibitem{Wang:2020elp}
E.~Wang, H.~S.~Li, W.~H.~Liang and E.~Oset,
Phys. Rev. D \textbf{103}, 054008 (2021).

\bibitem{Deineka:2021aeu}
O.~Deineka, I.~Danilkin and M.~Vanderhaeghen,
Phys. Lett. B \textbf{827}, 136982 (2022).

\bibitem{Wei:2021usz}
L.~L.~Wei, H.~S.~Li, E.~Wang, J.~J.~Xie, D.~M.~Li and Y.~X.~Li,
Phys. Rev. D \textbf{103}, 114013 (2021).

\bibitem{Prelovsek:2020eiw}
S.~Prelovsek, S.~Collins, D.~Mohler, M.~Padmanath and S.~Piemonte,
JHEP \textbf{06}, 035 (2021).

\bibitem{Liu:2009qhy}
X.~Liu, Z.~G.~Luo, Y.~R.~Liu and S.~L.~Zhu,
Eur. Phys. J. C \textbf{61}, 411-428 (2009).

\bibitem{Liu:2017mrh}
M.~Z.~Liu, D.~J.~Jia and D.~Y.~Chen,
Chin. Phys. C \textbf{41}, 053105 (2017).

\bibitem{Sakai:2017avl}
S.~Sakai, L.~Roca and E.~Oset,
Phys. Rev. D \textbf{96}, 054023 (2017).

\bibitem{LHCb:2021auc}
R.~Aaij \textit{et al.} [LHCb],
Nature Commun. \textbf{13}, 3351 (2022).

\bibitem{Li:2012ss}
N.~Li, Z.~F.~Sun, X.~Liu and S.~L.~Zhu,
Phys. Rev. D \textbf{88}, 114008 (2013).

\bibitem{Abreu:2022sra}
L.~M.~Abreu,
[arXiv:2206.01166 [hep-ph]].

\bibitem{Ohkoda:2011vj}
S.~Ohkoda, Y.~Yamaguchi, S.~Yasui, K.~Sudoh and A.~Hosaka,
Phys. Rev. D \textbf{86}, 014004 (2012).

\bibitem{Dai:2022ulk}
L.~R.~Dai, E.~Oset, A.~Feijoo, R.~Molina, L.~Roca, A.~M.~Torres and K.~P.~Khemchandani,
Phys. Rev. D \textbf{105}, 074017 (2022).


\bibitem{He:2015mja}
J.~He,
Phys. Rev. D \textbf{92}, 034004 (2015).

\bibitem{LHCb}
https://home.web.cern.ch/news/news/physics/lhcb-discovers-three-new-exotic-particles.







\end{thebibliography}
\end{document}